\documentclass[1p]{elsarticle}

\usepackage{hyperref}
\usepackage[utf8]{inputenc}
\usepackage[english]{babel}
\usepackage[T1]{fontenc}
\usepackage{amsmath}
\usepackage{amsfonts}
\usepackage{amssymb}
\usepackage{graphicx}
\usepackage{xcolor}
\usepackage{booktabs}

\journal{CIRP Journal of Manufacturing Science and Technology}

\begin{document}

\begin{frontmatter}


\title{Root Cause Analysis in Lithium-Ion Battery Production with FMEA-Based Large-Scale Bayesian Network }

\author[bmwaddress]{Michael Kirchhof\corref{mycorrespondingauthor}}
\cortext[mycorrespondingauthor]{Corresponding author.}
\ead{michael.kirchhof@tu-dortmund.de} 

\author[bmwaddress]{Klaus Haas}
\ead{klaus.haas@student.tugraz.at}

\author[bmwaddress]{Thomas Kornas}
\ead{thomas.kornas@bmw.de}

\author[braunschweig]{Sebastian Thiede}
\ead{s.thiede@tu-braunschweig.de}

\author[tugaddress]{Mario Hirz}
\ead{mario.hirz@tugraz.at}

\author[braunschweig]{Christoph Herrmann}
\ead{c.herrmann@tu-braunschweig.de}

\address[bmwaddress]{BMW Group, Technology Development, Prototyping Battery Cell, Lemgostrasse 7, 80935 Munich, Germany}

\address[tugaddress]{Graz University of Technology, Institute of Automotive Engineering, Inffeldgasse 11, 8010 Graz, Austria}

\address[braunschweig]{Technische Universität Braunschweig, Institute of Machine Tools and Production Technology (IWF), Universitätsplatz 2, 38106 Braunschweig, Germany}



\begin{abstract}
The production of lithium-ion battery cells is characterized by a high degree of complexity due to numerous cause-effect relationships between process characteristics. Knowledge about the multi-stage production is spread among several experts, rendering tasks as failure analysis challenging. In this paper, a new method is presented that includes expert knowledge acquisition in production ramp-up by combining Failure Mode and Effects Analysis (FMEA) with a Bayesian Network. Special algorithms are presented that help detect and resolve inconsistencies between the expert-provided parameters which are bound to occur when collecting knowledge from several process experts. We show the effectiveness of this holistic method by building up a large scale, cross-process Bayesian Failure Network in lithium-ion battery production and its application for root cause analysis.
\end{abstract}


\end{frontmatter}

\section{Introduction}

Given the necessity of CO2 reduction in the mobility sector, which is driven by the European Commission's upcoming new regulations for automotive manufacturers, the shift towards electrification can be observed as one major trend in the industry \cite{Pehnt2011}. Currently, lithium-ion battery (LIB) cells as energy carriers for electric vehicles are one key technology, due to their high energy density and long life cycles \cite{BRODD20132}. However, there are certain challenges yet to overcome. As of the current technical state of the art, LIB cell manufacturers face quality issues, which is reflected in scrap rates of $6$ to $12\%$ \cite{BRODD2013} \cite{VDMA}. This is not only a significant cost factor, but also affects the environmental impact, since LIB production accounts for $50\%$ of emissions during the production of an electric vehicle.

\begin{figure}[h]
\centering

\includegraphics[scale=0.5, page = 2, trim = 1cm 3cm 1cm 3cm,clip = true]{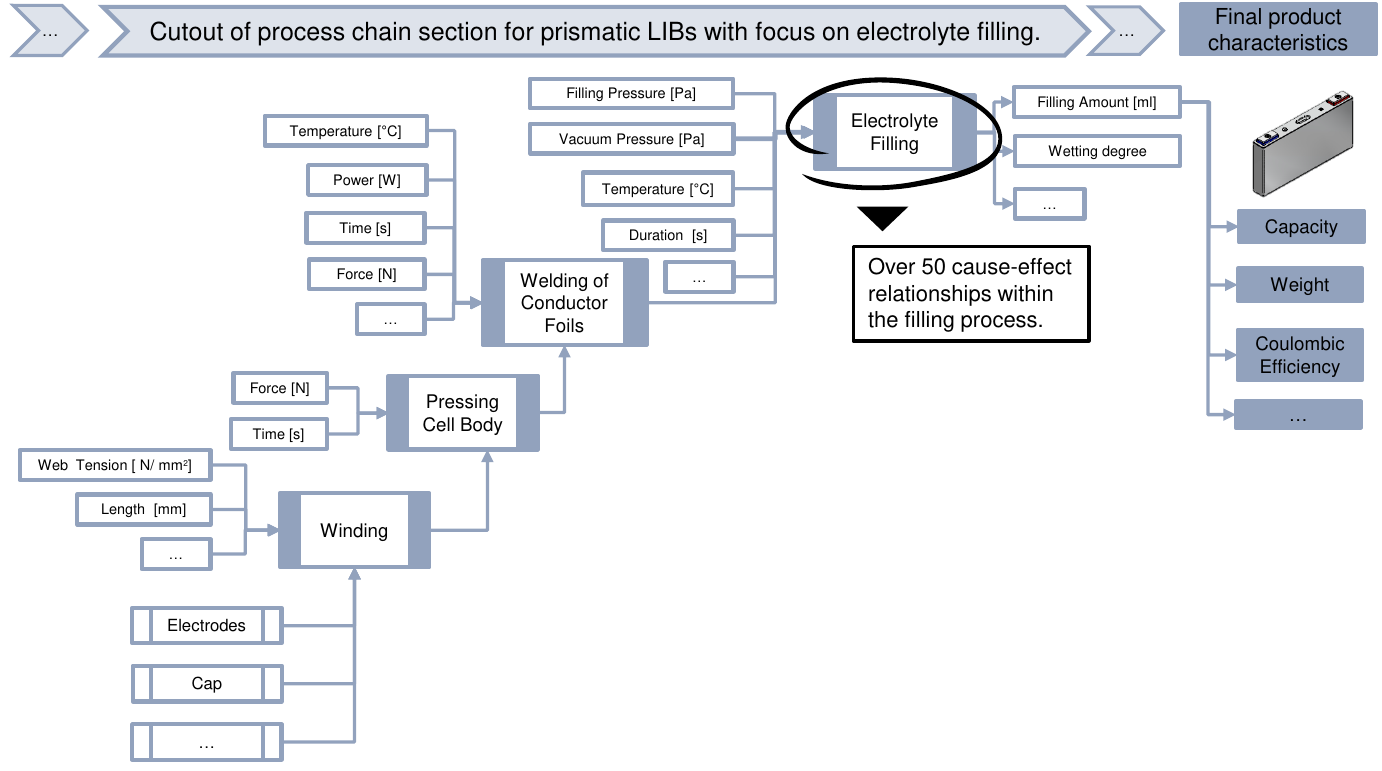}
\caption{Example of CERs in LIB cell production}
\label{fig:example}
\end{figure}

According to research, the scrap rates can be traced back to the high complexity in cell production as a result of many different process steps and a high amount of cause-effect-relationships (CERs) between process characteristics \cite{Westermeier2016} \cite{KORNAS1} \cite{KORNAS2} \cite{KORNAS3}. The production of LIBs involves about $600$ process characteristics such as machine parameters and other properties, whose CERs can be depicted as a network consisting of up to $2{,}100$ connections, $75\%$ of which are assumed to be essential for final cell quality \cite{Westermeier2016}. Figure \ref{fig:example} depicts some exemplary CERs in the "electrolyte filling" process.

Usually a lot of historical production data is available in series production enabling the application of various data-driven methods upon which failures can be detected and traced back to their roots \cite{Westermeier2016} \cite{Thiede1}. During prototyping, pilot series and ramp-up, the amount of available production data may still be low, so process corrections due to quality deviations and errors are carried out mostly based on expert knowledge  \citep{Westermeier2016} \citep{KORNAS2}. Considering the complexity, the utilization of expert knowledge for root cause analysis (RCA) gathered by the conduction of quality methods such as Failure Mode and Effect Analysis (FMEA) may easily become strenuous and time-consuming. Furthermore, inconsistencies and contradictions between different ratings can occur during the knowledge acquisition. This is because experts for individual process steps are not able to consider all interactions of their ratings across other process steps. Yet expert knowledge-based quality methods are essential, especially during ramp-up \cite{VDMA} \cite{KORNAS2}.

This paper presents an innovative, quality-oriented approach to creating and continuously improving RCA from expert knowledge in the complex process chain of early-stage LIB production by combining the benefits of a process FMEA failure network with those of a Bayesian Network. It is assumed that such an approach could reduce the time needed to identify root causes of detected failures and thus also help to improve overall production ramp-up time. In section \ref{chap:research}, the state of the art and research regarding methods of quality management is reviewed. Also, current applications of Bayesian Networks in quality assurance are presented. Afterwards, the methods for converting an FMEA into a valid Bayesian Network as well its possibilities for RCA are presented (\ref{chap:methods}). In section \ref{chap:application}, the method is applied to build a Bayesian Network for a RCA in LIB production, followed by a summary and prospects for further research (\ref{chap:conclusion}).

\section{State of the Art and Research} \label{chap:research}
\subsection{Quality Management and Cause-Effect-Relationships}

A production ramp-up entails special requirements to applied quality methods that consider CERs since data availability is low and parameter specification limits are often not fully defined. Therefore, traditional approaches such as Statistical Process Control (SPC), which are used to optimize processes by means of statistical methods, are not applicable. This limits the selection of applicable quality assurance methods primarily to those that do not rely on quantitative data \citep{KORNAS2}, but rather on expert knowledge. Methods that are based on expert knowledge are henceforth referred to as expert-based methods.

Various expert-based methods designed for the identification and analysis of CERs are available, although only few are suitable for complex manufacturing process chains with a high amount of CERs. Fault Tree Analysis (FTA) and Failure Mode and Effects Analysis (FMEA) are considered to be the most well-established among these \citep{Cristea.2017}.

FTA generally follows the principle of building top-down fault trees with numerical information about the failure occurrences, which can be linked to logical gates \citep{Bruggemann.2012}. The method was designed to analyze malfunctions of subsystems within a larger technical system. Another quality management tool that is often applied to CERs is the Ishikawa Diagram. However, this tool is solely a graphical way to depict CERs without yielding an actual underlying analysis \citep{Benes.2011}.

FMEA comprises an expert-based analysis framework for risk and failure prevention in technical domains with analogies to FTA \citep{Werdich.2012} \citep{Bruggemann.2012}. The FMEA, unlike FTA, also contains qualitative information about the failures, such as correctional measures and failure severity estimations. Different derivatives have emerged from research and industrial projects \cite{Spreafico.2017} since the initial introduction of FMEA in 1949, while VDA 4.2 (acronym for the German Association of the Automotive Industry) \citep{VDA42} distinguishes between the following main types: Process FMEA (PFMEA) and Design FMEA (DFMEA). DFMEA aims at analyzing the product design itself in terms of quality-critical aspects, while PFMEA was developed to investigate manufacturing or assembly processes and the potential failure CERs involved in these \citep{Kahrobaee.2011} \cite{renu2016}. Other than FMEA or FTA, which are both based on CERs of failures, the method presented in \citep{Westermeier2016} provides a framework for the expert-based assessment of CERs of process characteristics without breaking it down into potential failures of these. Since a process characteristic may have multiple potential failures subordinated to it, the level of detail in this method is insufficient for an RCA of failures and deviations in the complex LIB production. RCA is a term in quality management that generally refers to the reactive identification process of a failure's root cause, wherein knowledge that has been gained during application of other quality methods, such as FMEA, can be utilized \cite{Vanden.2008} \cite{Wilson.1993}.
 
FTA and FMEA both can be seen as directed acyclic graphs (DAG) \citep{Lenz.2004} \citep{Katoen.2017}. Therefore, both methods can be utilized as the starting point of an RCA. However, the qualitative information in an FMEA also allows for a preventive failure consequence assessment without requiring much more effort than the creation of the plain failure network with its occurrence rates. This may add further value to the overall process quality. 

\subsection{Application of Bayesian Networks in Root Cause Analysis}
The cognitive effort of manually conducting an RCA solely based on FMEA failure nets would increase with the complexity of the network and the amount of CERs involved \cite{Halford.2005}. Various approaches have tried to resolve this shortcoming by making use of different statistical models, each with its own advantages. Extending failure nets to Bayesian Networks has so far shown to be a promising concept \cite{Spreafico.2017}. While Bayesian Networks demonstrate a notable amount of robustness against deviations in their parameters \citep{Henrion1996}, they nevertheless lack the direct modeling of uncertainty of expert statements. Other potential systems for root cause analysis, such as Credal Networks \citep{ANTONUCCI2009} or Bayesian Models \citep{SEIXAS2014}, provide a framework to explicitly incorporate uncertainties. This, however, comes at the cost of higher computational complexity especially for large-scale models. Even medium-sized Credal Networks are reported to vastly exceed real time inference \citep{IDE2008}. As the size of the model built in this work is one order of magnitude bigger, Bayesian Networks are preferred due to their scalable inference algorithms.



Existing approaches to create Bayesian Networks either rely on quantitative production data \cite{Khakzad.2011} \cite{Lokrantz.2018} or they have been designed specifically for simple process chains, and thus do not involve a proper knowledge acquisition procedure that could be carried out in a reasonable amount of time for complex process chains \citep{mcnaught2011} \cite{Lee.2001} \citep{Huang.2008}. Additionally, none of these provides a framework for the continuous improvement of the knowledge base
. Still, a Bayesian Network bears intrinsic potential since it can improve itself further and consequently also the FMEA by learning from failures that have occurred after its initial creation. In the course of the process chain's ongoing growth and maturation, this information can either be collected in the form of error protocols \citep{renu2016} or from user interactions when inquiring the network for the root cause of a present failure \citep{mcnaught2011}.

Since LIB cell production consists of interdisciplinary process steps with different process experts in charge, another challenge arises: Mathematical inconsistencies between inter-process expert ratings are bound to occur. In order to ensure that the knowledge acquisition is carried out in a reasonable time while full consistency without contradictions is maintained, experts would need to consider all other ratings that have been made before. Since the human ability to consider multiple interdependencies at a time is limited \cite{Halford.2005}, a validation algorithm is needed to support this process.


\subsection{Research Demand}
To summarize the current state of research, there is no holistic expert-based method that enables a combined way of creating structured knowledge along a complex manufacturing process chain that can subsequently be used for an RCA. The idea of extending a failure network based on expert ratings to a Bayesian Network, although not new, is highly promising. It would provide an opportunity for probabilistic fault diagnosis even in cases in which no measurement data is available. Failure information, which is gathered as production advances, can be used to improve the network. Still, existing approaches do not provide sufficient ways for knowledge acquisition in complex processes. Considering hundreds of CERs in a process chain, unstructured knowledge acquisition for uncertain reasoning would be too demanding and could lead to contradicting information. In the following chapter, a new approach to overcome these shortcomings is presented.


\section{Methods} \label{chap:methods}

Based on the current state of research, a new method was developed in order to fill the research gap for expert-based RCA in complex manufacturing process chains. A failure network from an FMEA in the battery manufacturing process chain is used as a basis to build a Bayesian Network, that can be utilized for an RCA. Due to the advantages that FMEA has over FTA (as described in section \ref{chap:research}) an FMEA was chosen to represent the basis for the Bayesian Network.

Leaky Noisy-OR Gates are then used to reduce the number of probabilistic estimates that are needed to create a Bayesian Network and an evolutionary algorithm maintains consistency during knowledge acquisition. After the introduction of the statistical methods, the knowledge acquisition process followed by the RCA is explained in the following sections.

\subsection{Bayesian Networks}
From a statistical perspective, each failure surveyed in the FMEA is represented by a random variable with binary outcomes - either the failure has occurred or it hasn't. When we denote the number of failures in the FMEA as $n$, we can write all these variables inside a random vector $X = (X_1, \dotsc, X_n)$. To represent all influences between the failures, the joint distribution $P(X)$ describing the statistical relations between all variables has to be found. This is a difficult task when many variables are involved, so it requires a way to structure and simplify $P(X)$. 

\begin{figure}[h]
\centering

\includegraphics[scale=0.5, page = 4, trim = 1.2cm 7cm 1cm 4.2cm,clip = true]{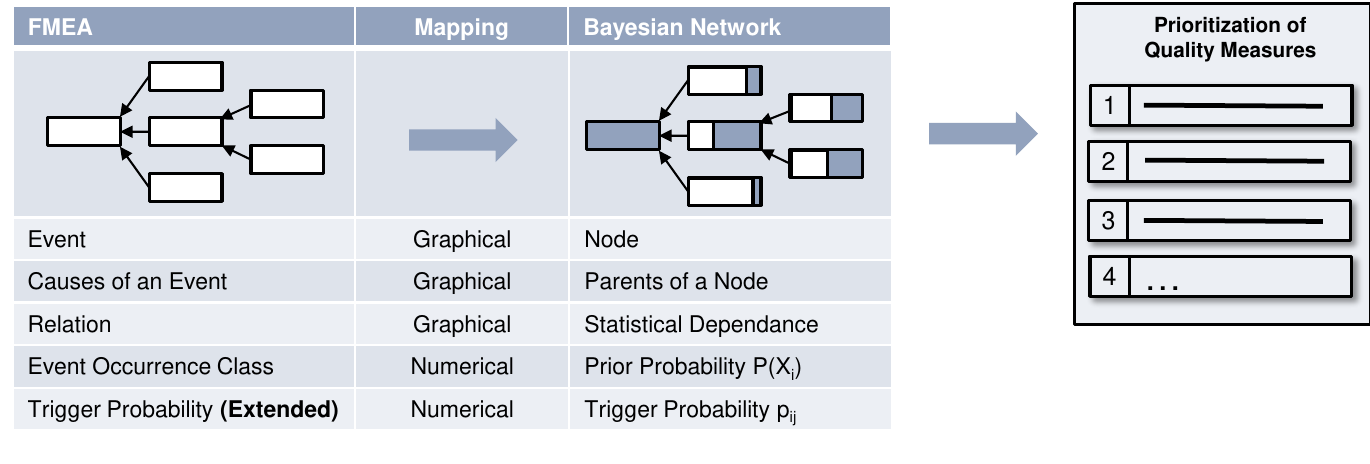}
\vspace{-0.5cm}
\caption{Translation of an FMEA into a Bayesian Network.}
\label{fig:comparison}
\end{figure}

The key idea behind a Bayesian Network is to assume that the probability distribution of each variable depends only on a subset of other variables, its so called \textit{parents} $\text{Pa}(X_i) = (X^{(i)}_1, \dotsc, X^{(i)}_J)$. Given these variables, the local distribution $P(X_i\,|\,\text{Pa}(X_i))$ is specified in the form of a conditional probability table. It shows the probability that $X_i$ occurred for each of the combinations of the parents' states. If a variable does not have any parents, it is called a \textit{root node} and solely the probability of $X_i$ occurring is needed, which is referred to as its \textit{prior}. Once all of these local distributions are specified, they can be multiplied to obtain the joint distribution $P(X)$. In terms of graphic representation, the variables stand for nodes in a graph. Arcs exist between a node and its parents, showing their statistical dependency. An in-depth description of Bayesian Networks is found in \cite{heckerman1995}. Figure \ref{fig:comparison}. The trigger probabilities mentioned in the figure will be explained in section \ref{chap:leakynor}, but are shown here for the sake of completeness.

Various forms of inference can be made on the completely specified network. Given information about the observed state of some failures, the evidence, the a-posteriori probabilities of all other failures are calculated. One particular use case of these predictions is RCA. This method is applied when a specific failure has occurred and the goal is to identify cause of this failure, possibly given some more information about other failures. An example of this is given in section \ref{chap:RCA}.

The probabilities needed in those inferences can be calculated exactly or, to avoid the runtime-heavy computations, approximated or simulated. Due to the size of our failure network, we decide for simulations using the likelihood weighting algorithm \citep{shwe1991}. Roughly speaking, this algorithm randomly generates a certain number of artificial observations from the network. Every failure that has evidence is forced to take its known evidence state. In a last step, the observations are weighted according to the conditional probability of their evidence to gain a non-biased result. The number of generated observations can be increased for more accurate results at the cost of longer computing times. Here, $10^5$ observations have shown to be a good compromise.

\subsection{Leaky Noisy--OR Gate} \label{chap:leakynor}
As described above, the conditional probability of each variable given an arbitrary combination of its parents' states has to be specified when building a discrete Bayesian Network. However, this number of probabilities rises exponentially in the number of its parents. For example, if a variable has $10$ parents, there are $2^{10} = 1024$ different combinations of parent states, and the probability of the variable occurring or not occurring has to be specified for each of these. \cite{SRINIVAS1993} points out that when asking experts for that many estimates, the quality of the given estimates may decrease. To counteract this, a parametrized distribution can be used to calculate the conditional probability tables from less inputs given certain assumptions.

A common choice for such a distribution is the \textit{Noisy-OR Gate (N-OR)} \citep{pearl1988}. It makes it possible to generate the whole conditional probability table by supplying only one probability per variable, thereby reducing the exponential number of parameters to a linear one. The additional parameter is called the \textit{trigger probability} $p^{(i)}_j = P(X_i = 1 \,|\, X^{(i)}_1 = 0, \dotsc, X^{(i)}_j = 1, \dotsc, X^{(i)}_n = 0)$. It gives the probability that $X_i$ is active given that only one of its parents $X^{(i)}_j$ is active; or in the context of FMEA: The probability that a failure $X^{(i)}_j$ will trigger the failure $X_i$, given no other known or unknown failures occurred. We use Diez's parameterization of Noisy-OR as the research suggests it provides the best results when surveying trigger probabilities from experts \citep{zagorecki2004}.

One of the assumptions of a N-OR is that there are no other causes for $X_i$ than its parents \cite{russell2016}, which are the causes surveyed from the process experts. However, it would be naive to assume that there are no other possible causes besides these. Therefore, a \textit{leak variable $L^{(i)}$} is introduced \citep{diez1993}. It represents all unknown causes of a failure and can be thought of as an additional parent with a trigger probability of $1$. To calculate the prior probability of this variable, the gap between the prior probability of $X_i$ surveyed within the FMEA and the marginal probability of $X_i$ given all known causes can be utilized. The exact formula and its proof are found in the appendix.


\subsection{Latent Variables for Aggregation}
Using N-OR solves the problem of the exponential rise in the required probabilities when surveying the domain experts. However, when doing inference on the network, all possible combinations of parent states still have to be considered and thus the runtime complexity is still exponential in the number of parents. To solve this issue, latent variables that aggregate several parents can be inserted between the parents and the children. This does not affect the conditional probability table of the child as N-OR belongs to the family of decomposable distributions \citep{heckerman1996}. Yet it offers scalability and decreases the inference time in large networks \citep{heckerman1996}, enabling its use in multi-stage production settings. 

\begin{figure}[h]
\centering

\includegraphics[scale=0.5, page = 6, trim = 1cm 3cm 1cm 3cm,clip = true]{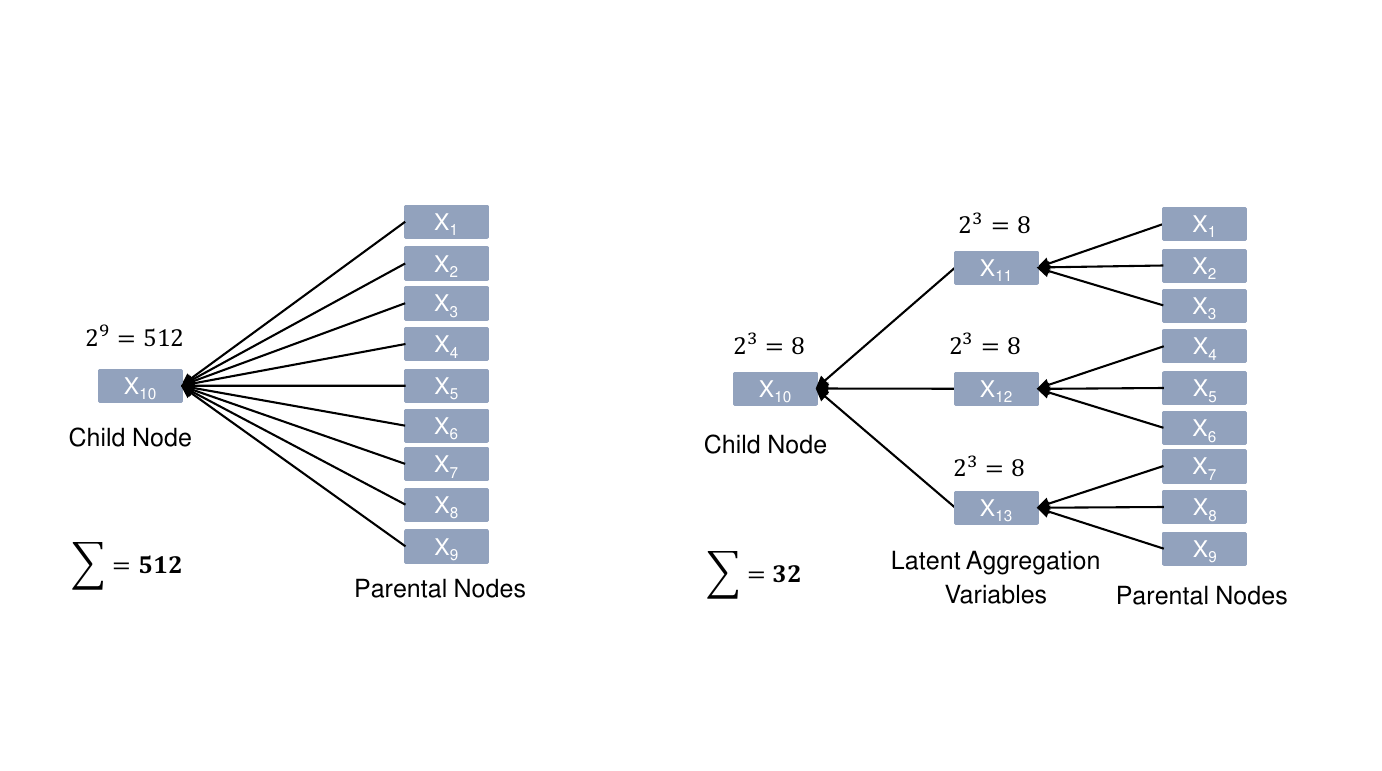}
\vspace{-1cm}
\vspace{-0.5cm}
\caption{Bayesian Network without (left) and with (right) aggregation variables. The numbers show the storage required for the conditional probability tables.}
\label{fig:netGroups}
\end{figure}

Figure \ref{fig:netGroups} shows an example of a child with $9$ parents. By building groups of three, the number of probabilities required to make computations on the network can be reduced from $2^9=512$ (left) to $4 \cdot 2^3 = 32$ (right). Alternatively, it could also be split recursively into groups of two, again resulting in $8 \cdot 2^2 = 32$ probabilities. Unfortunately, there is no research on how the group size affects the variance during likelihood weighting simulations. Theoretically speaking, a smaller group size and thus more simulated failures might lead to higher variance. Thus, in this work the group size is chosen to be as low as possible to ensure smooth operation while not introducing too many latent aggregation variables, resulting in a maximum group size of $5$.

\subsection{Recommending Consistent Networks} \label{chap:recommend}
The fact that FMEA surveys prior probabilities even for intermediate failures makes it possible to check for so called inconsistencies: A failure might happen to be over-explained by its causes, meaning that given the prior and trigger probabilities of the causes, the failure should occur more often than the experts expect. Formally, this means that the marginal probability of a variable given all its parents is higher than its specified prior probability. This occurs due to a mismatch in the variable's prior probability and its parents' prior and trigger probabilities. Note that this comparison is also dependent on the model choice as the marginal probability is calculated using the model. Consequently, the following procedure is only applicable to Bayesian Networks with N-OR assumption and needs to be altered if other models are used.

As will be shown in section \ref{chap:application}, there can be several interconnected inconsistencies within a network. To support the expert in resolving these, an algorithm has been developed that searches for a consistent network that is as close to the expert-provided failure network as possible. This suggestion is presented to the expert together with their own FMEA network to help remove the inconsistencies.

The optimization algorithm will search for consistent prior probabilities and trigger probabilities. However, in FMEA the expert does not directly give prior probabilities, but only occurrence rate classes. Thus, for the prior probabilities, we have to measure the distance of a suggested network to the expert network in the space of these occurrence rate classes. Table \ref{tab:occRate} shows the probability intervals associated with each occurrence rate class based on the suggestions of \cite{VDA42}.

\begin{table}[hb]
\caption{Prior probabilities associated with each occurrence rate class}
\label{tab:occRate}
\center
\begin{tabular}{rrr}
 \toprule
	FMEA Occurrence Rate & \multicolumn{2}{r}{Probability Interval} \\
 \midrule
	$1$ & $ [ 0,$ & $ 1 \cdot 10^{-6} ] $ \\
	$2$ & $ (1 \cdot 10^{-6} ,$ & $ 50 \cdot 10^{-6}] $ \\
	$3$ & $ (50 \cdot 10^{-6},$ & $ 100 \cdot 10^{-6}] $ \\
	$4$ & $ (100 \cdot 10^{-6},$ & $ 1 \cdot 10^{-3}] $ \\
	$5$ & $ (1 \cdot 10^{-3},$ & $ 2 \cdot 10^{-3}] $ \\
	$6$ & $ (2 \cdot 10^{-3},$ & $ 5 \cdot 10^{-3}] $ \\
	$7$ & $ (5 \cdot 10^{-3},$ & $ 10 \cdot 10^{-3}] $ \\
	$8$ & $ (10 \cdot 10^{-3},$ & $ 20 \cdot 10^{-3}] $ \\
	$9$ & $ (20 \cdot 10^{-3},$ & $ 50 \cdot 10^{-3}] $ \\
	$10$ & $ (50 \cdot 10^{-3},$ & $ 1] $ \\
 \bottomrule
\end{tabular}

\end{table}

The problem of searching the closest consistent network can be formulated as a constrained optimization problem with quadratic loss:
\begin{gather*}
	\text{argmin}_p\, {\left\lVert  \frac{c}{\left\lVert c\right\rVert} \cdot (q(p) - q_e) \right\rVert}^2 \\
	\text{constraint: the network generated by p has no inconsistencies.} 
\end{gather*}
Here, $p$ is a vector containing the prior and trigger probabilities of a suggested network. As explained above, the vector $q(p)$ contains the corresponding occurrence rate classes and trigger probabilities. $q_e$ contains the expert-given parameters. The vector $c$ contains scalars that represent the costs to change the individual parameters. By utilizing $c$, different scales between occurrence rates (1 to 10) and trigger probabilities (0 to 1) can be taken into account. Moreover, $c$ could be used to represent the uncertainty of expert ratings. Parameters the expert is not sure about can be changed at lower costs than high-confidence parameters. 

The constraint in the above optimization formula can be broken down into several smaller constraints. A network has no inconsistencies if, and only if, the marginal probabilities are smaller than the priors for all variables. This way, the consistency of each variable becomes an individual constraint. Unfortunately, these marginals and their derivatives have no handy functional form, making the optimization infeasible. To handle this issue, the constraints are considered differently: Instead of forcing all suggestions to adhere to all constraints, the number of violated constraints $n_\text{incon}(p)$ is added as penalty factor. A hyperparameter $\alpha$ is introduced to balance resolving the highest possible number of inconsistencies and staying close to the expert estimate. Finally, we arrive at the following formula:
\begin{align*}
	\text{argmin}_p \, {\left\lVert  \frac{c}{\left\lVert c \right\rVert} \cdot (q(p) - q_e)\right\rVert}^2 + \alpha \cdot n_\text{incon}(p)\,\,.
\end{align*}

Due to the rough form of this optimization formula, we apply a genetic algorithm \citep{eiben2003} to search for the optimal parameter vector $p$. Several customizations are made to take advantage of the network structure. When crossing over two suggestions, the parameters are first bundled by the variable they belong to (with trigger probabilities belonging to the variable they trigger) before conducting a uniform cross-over. During mutation, we use a uniform distribution to select a mutation shift for each parameter. For prior probabilities, the span of this distribution equals the width of the probability interval of the corresponding occurrence class in both the negative and the positive directions. Trigger probabilities are not allowed to grow above their expert-given value as increasing a trigger probability beyond this limit can never resolve an existing inconsistency while it always increases the distance to the original expert suggestion, resulting in a worse suggestion. Moreover, the probability of mutation is adapted depending on whether the population's best suggestion has enhanced or become stuck during the previous iteration.

\subsection{Building a Bayesian Network out of an FMEA} \label{chap:BuildingNet}
\begin{figure}[h]
\centering
\includegraphics[scale=0.35]{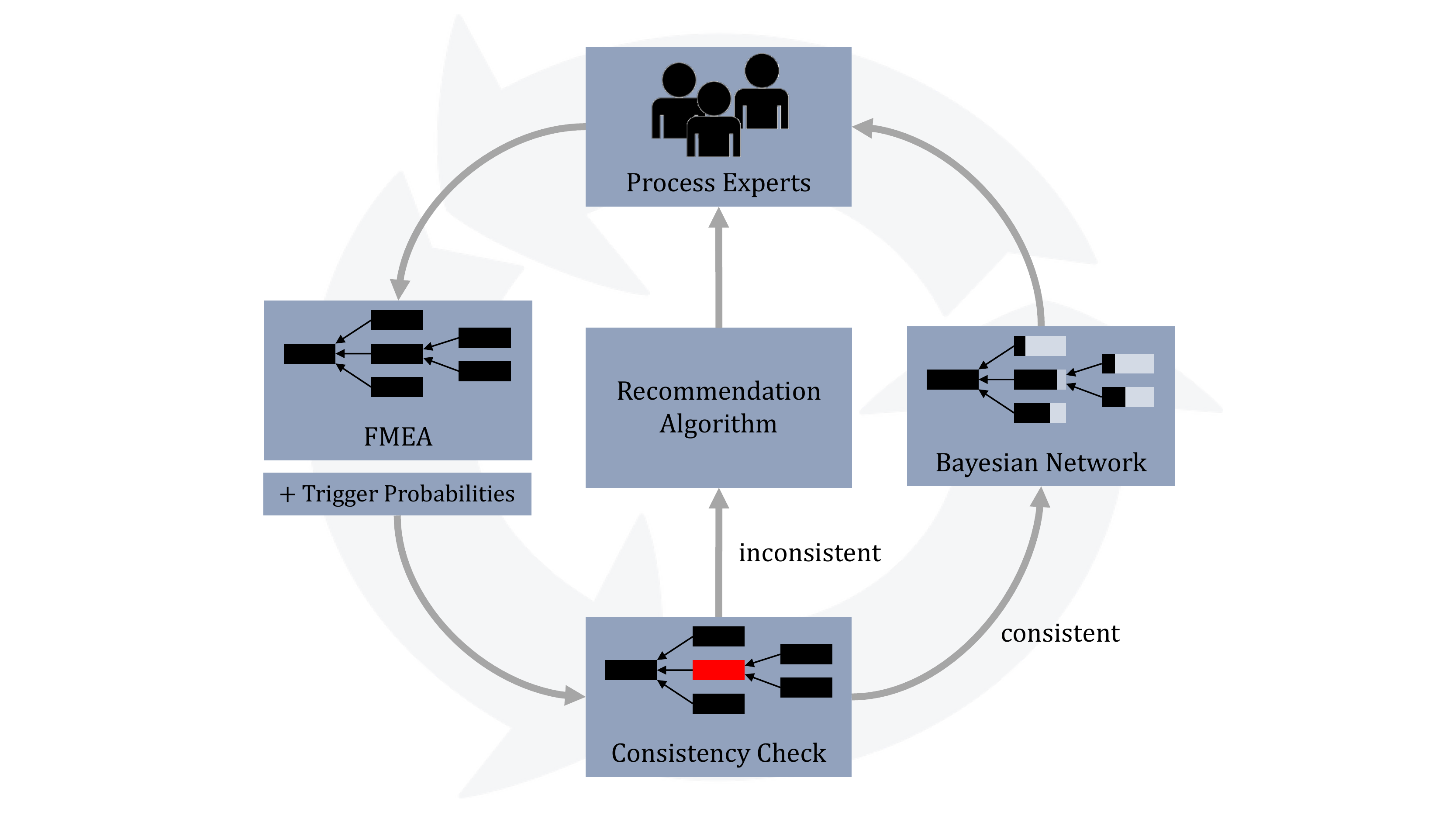}
\caption{Process for creating a Bayesian Network from expert knowledge}
\label{fig:overview}
\end{figure}

The whole processes of creating a Bayesian Network from expert knowledge described in the last sections is summarized in this section and visualized in Figure \ref{fig:overview}. In order to build a Bayesian Network, a proper knowledge base needs to be prepared. This is done by conducting the FMEA in the examined process chain by common FMEA procedure \citep{Werdich.2012}. The procedures may slightly vary according to country or industry, so it is suggested to select one according to the individual requirements of the relevant company. When carrying out the FMEA, experts identify failures throughout the process chain and then try to graphically depict their CERs, which ultimately results in the failure net. After that, experts need to conduct the actual rating of the identified failure CERs in terms of their severity, probability of occurrence and detectability. 
Along with the occurrence rates, experts are surveyed about the abovementioned trigger probabilities. 

During the FMEA, each process step is analyzed in chronological order. In order to mitigate the risk of inconsistencies, the present failure network with its prior and trigger probabilities is checked for consistency as outlined in section \ref{chap:recommend}. Possible lacks in understanding can be revealed by high leak probabilities, and inconsistencies are resolved by the expert with the help of the aforementioned algorithm. Once the FMEA is completed and all trigger probabilities attached accordingly, the full Bayesian Network is specified, which can be used for inferential inquiries. It can be continuously updated with new failures or other process knowledge. 

\subsection{Implementation of the Root Cause Analysis} \label{chap:RCA}

Once the engineer or expert has observed a failure in the process chain whose root cause could not be instantly determined, an RCA according to this method can be applied. The user can utilize the Bayesian Network to figure out the most likely reasons for the occurrence of the present failure.

\begin{figure}[ht]
\centering
\includegraphics[scale=0.5, page = 5, trim = 1cm 3cm 3cm 3cm,clip = true]{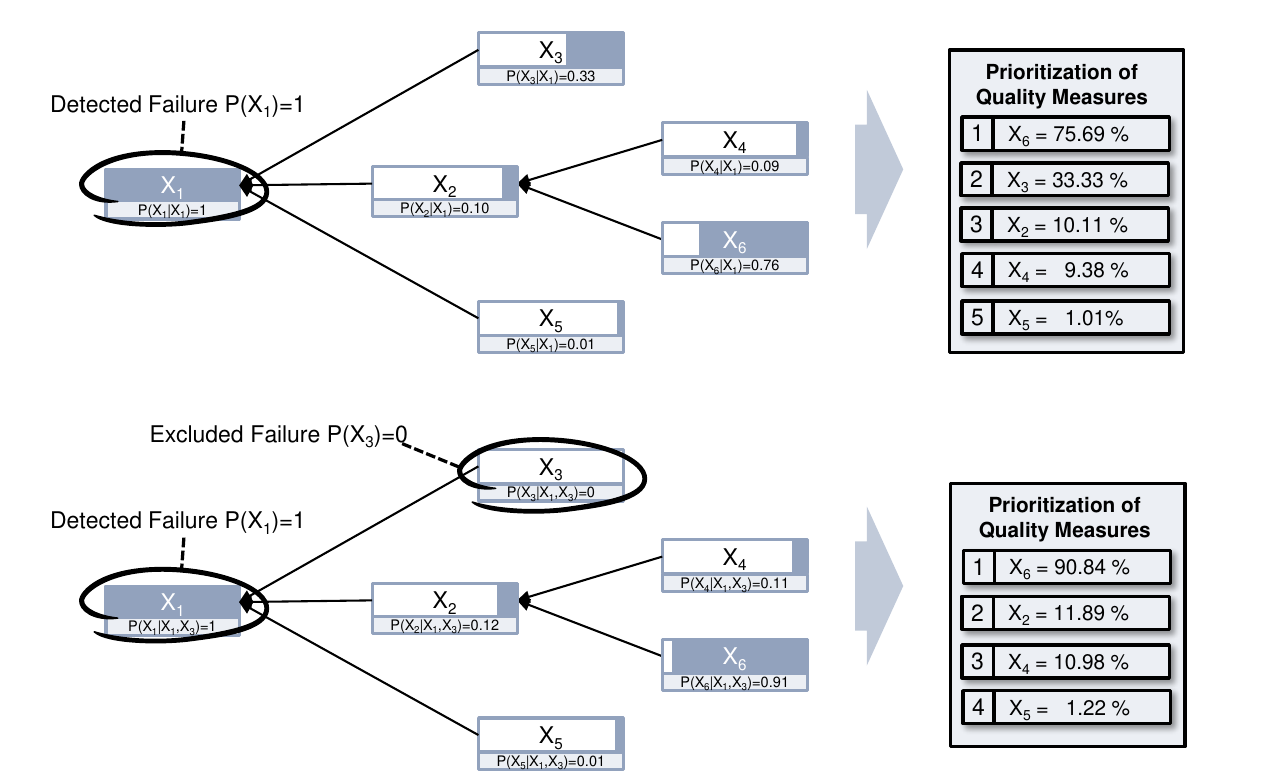}
\caption{Bayesian Networks with evidence in node $X_1$ (top) and nodes $X_1$ and $X_3$ (bottom).}
\label{fig:netEvidence}
\end{figure}

A minor example of this inference process is shown in Figure \ref{fig:netEvidence}. Here, an exemplary failure network with six failures is given. $X_1$ is the failure that initially occurred -- thus, its probability is set to $1$ -- and whose reason is to be found (top). By using the probabilities returned by the network, experts decide to verify failure $X_3$. They discover that $X_3$ did not occur, and feed this back into the network by setting its probability to $0$ (bottom). The network now shows that given this additional evidence, $X_6$ is the most likely reason for $X_1$ with a probability of occurrence of $90.84\%$. This interactive process is iterated until the root cause is found. A possible result of such an inference may be that $X_6$ has occurred and caused $X_2$ to happen, which in return triggered $X_1$. 

\section{Application} \label{chap:application}
The previously described method was applied in the lithium-ion battery prototyping production at BMW Group in Munich, Germany. In the following sections, we describe the derived failure network, the challenges faced regarding inconsistencies, as well as the implementation of the root cause analysis tool.

\subsection{Network and Inconsistencies}
The network surveyed through FMEA consists of $432$ failures and $1{,}098$ CERs between them. $219$ failures do not have any incoming CER, meaning that no cause is known so far. There are, however, failures with up to $32$ incoming CERs, underlining the importance of latent nodes for aggregation. $37\%$ of all CERs connect failures across different process steps. 

\begin{figure}[h]
\centering
\includegraphics[width = \textwidth]{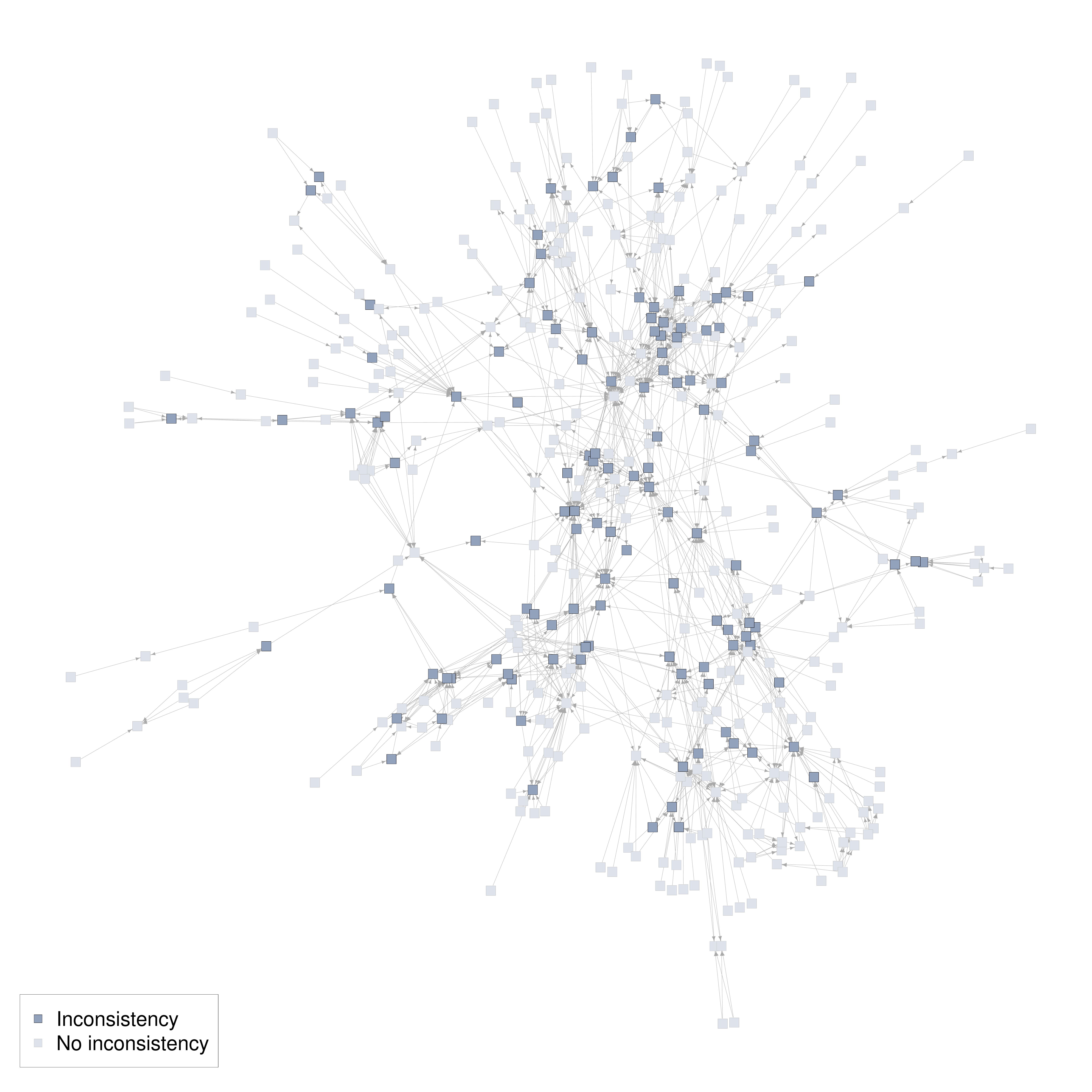}
\vspace{-1cm}
\caption{Initial failure network with highlighted inconsistencies}
\label{fig:inconsistencies}
\end{figure}

\begin{figure}[h]
\centering
\includegraphics[scale = 0.5]{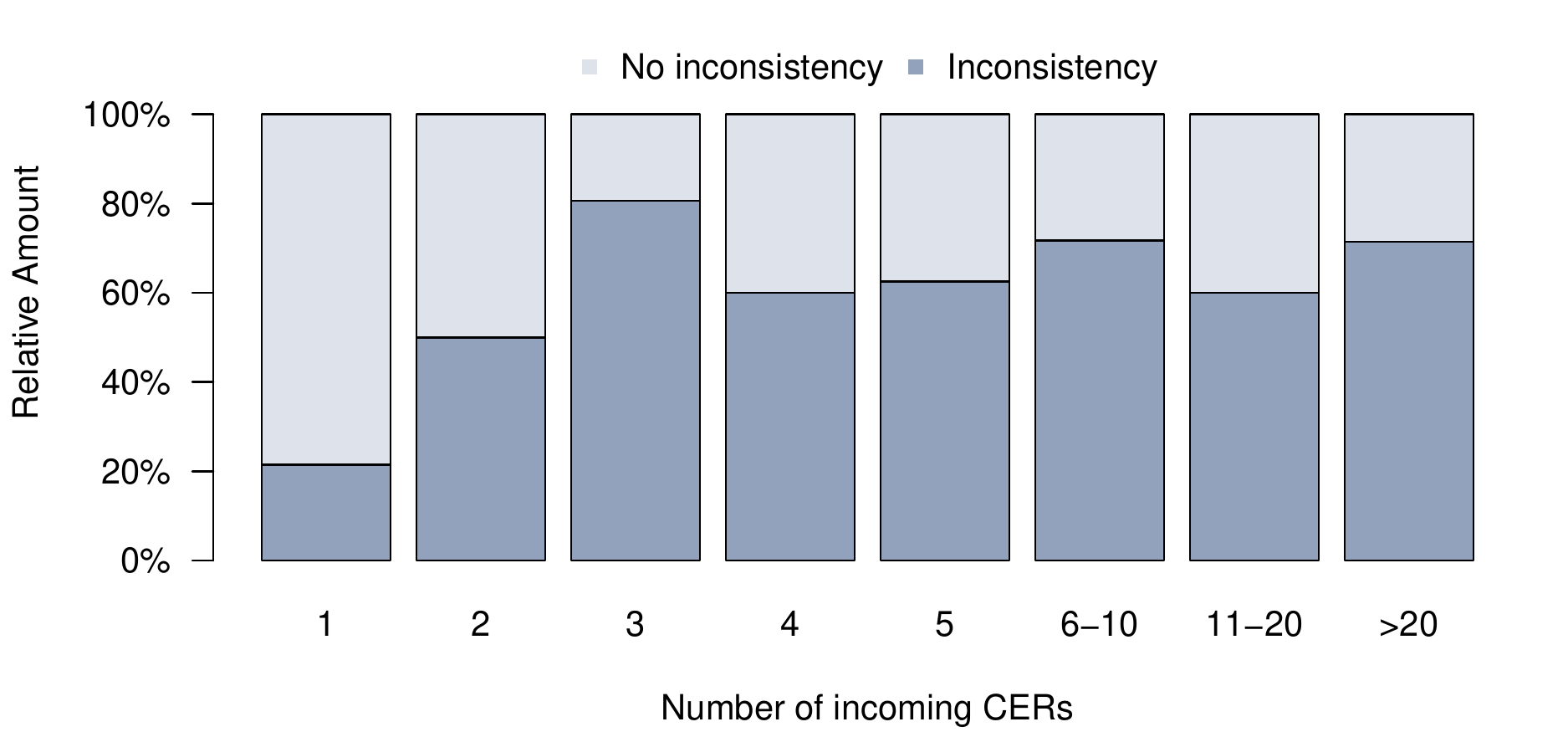}
\vspace{-0.2cm}
\caption{Percentage of failures that show inconsistencies grouped by the number of incoming CERs of the corresponding failures (failures without incoming CERs excluded).}
\label{fig:inconsistenciesCER}
\end{figure}

Inconsistencies occurred in $121$ of the $213$ failures with incoming CERs. Figure \ref{fig:inconsistencies} visualizes that these are spread throughout the whole network. In Figure \ref{fig:inconsistenciesCER}, the relative amount of inconsistencies depending on the number of incoming CERs is displayed in the vertical layer. It can be seen that even failures with only one incoming CER happen to be inconsistent. Failures with $3$ or more incoming CERs are considerably more often inconsistent. The number of inconsistencies does not further increase beyond $3$ incoming CERs.

In order to handle the inconsistencies, the prior and trigger probabilities were revised $15$ times with the aid of the recommendation algorithm. One process step showed no inconsistencies in its initial expert given parameters and thus did not need to be revised. In contrast, in some process steps that included inconsistencies, experts rejected parts of the computer-aided recommendations, so that multiple revisions and recommendations were necessary to reach a satisfactory solution. 

\begin{figure}[h]
\centering
\includegraphics[width = \textwidth]{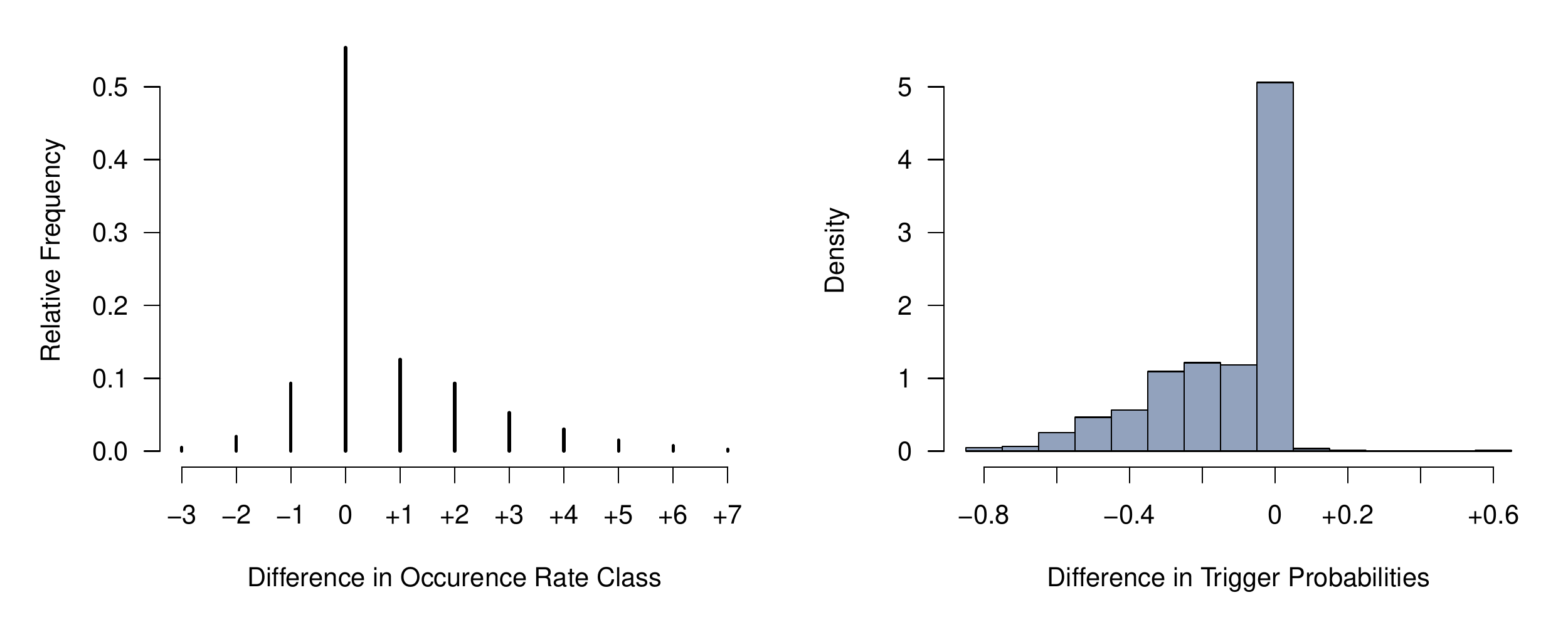}
\vspace{-1cm}
\caption{Comparison of the parameters in the initial and the final, consistent network.}
\label{fig:differences}
\end{figure}

The difference between the initial, inconsistent network and the final, consistent network is portrayed in Figure \ref{fig:differences}. The majority of both trigger probabilities and occurrence rate classes remained unchanged. Only few trigger probabilities were increased, which happened during the revisions carried out by the experts. The fact that almost all changed trigger probabilities decreased may be due to the suggestion algorithm, which was not permitted to increase them over their initial values, as explained in section \ref{chap:recommend}. 

\begin{figure}[h]
\centering
\includegraphics[width = 0.5 \textwidth]{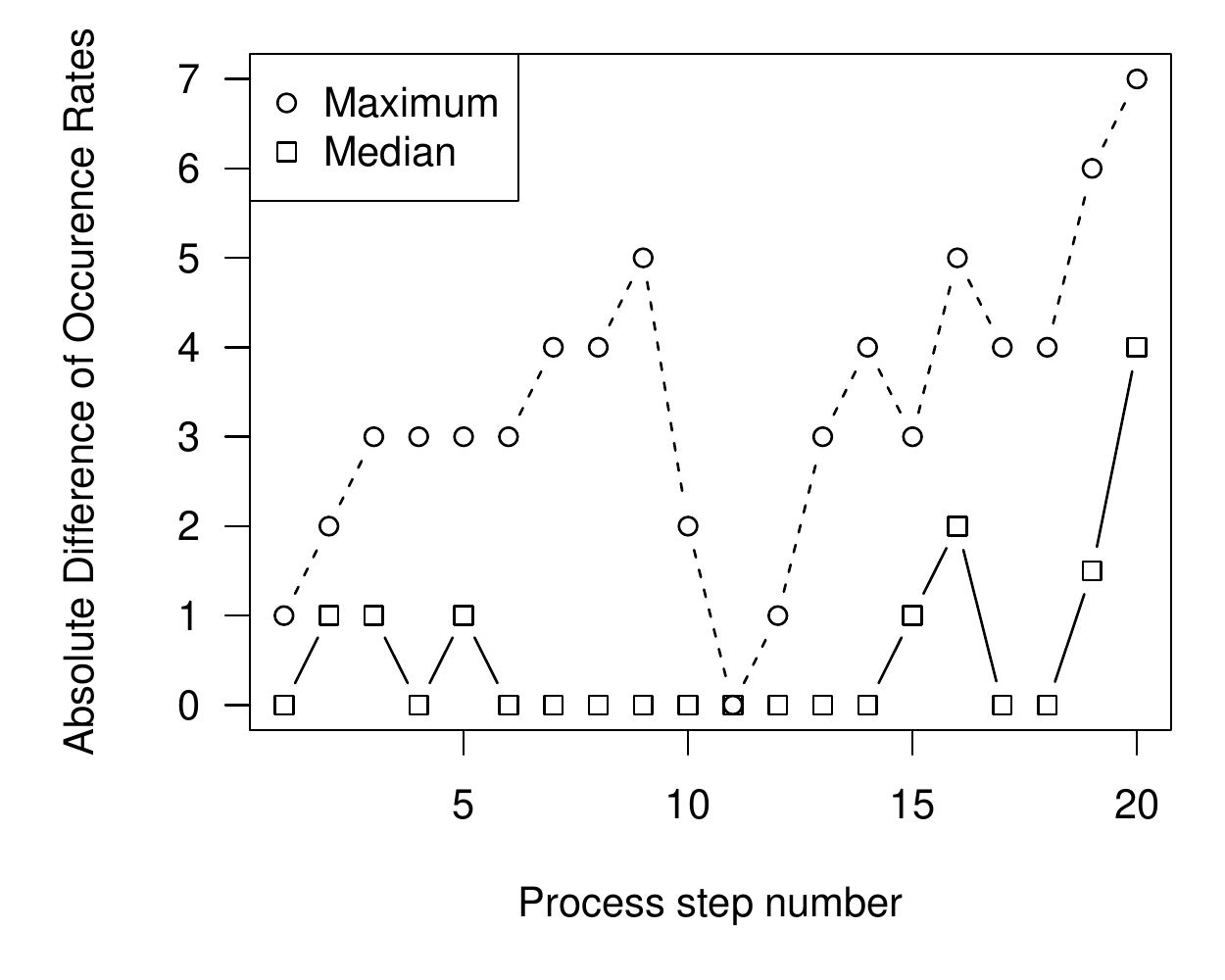}
\vspace{-0.5cm}
\caption{Differences of FMEA occurrence rates between the initial and the final, consistent network per process step.}
\label{fig:differencesPerProcess}
\end{figure}

The changes in the occurrence rate classes should be interpreted carefully as these classes follow an ordinal and not a metric scale. When changes were applied to the occurrence rates, these were increased for the most part. In Figure \ref{fig:differencesPerProcess}, the absolute value of these changes is presented per process step in their chronological order. A slight tendency for higher changes in later process steps can be seen, albeit ideally, the extent of changes should be similar for all process steps. However, to exclude the possibility of this being a random effect and to make well-founded statements on the extend and reasons for this trend, further experiments are required.

\subsection{Implementation and User Interface}

On the software side, APIS IQ-FMEA-L \cite{APIS2018} was used to conduct the FMEA. Once exported, the failure network was transmitted to an R \cite{R2019} script that executed the suggestion algorithm using the package GA \cite{Scrucca2013}. The consistent network was transformed into a Bayesian Network using the package bnlearn \cite{Scutari2010} to perform the RCA. The user interface for root-cause analysis was deployed on a server as an R-Shiny \cite{Chang2019} application.

\begin{figure}[h]
\centering
\includegraphics[width = \textwidth]{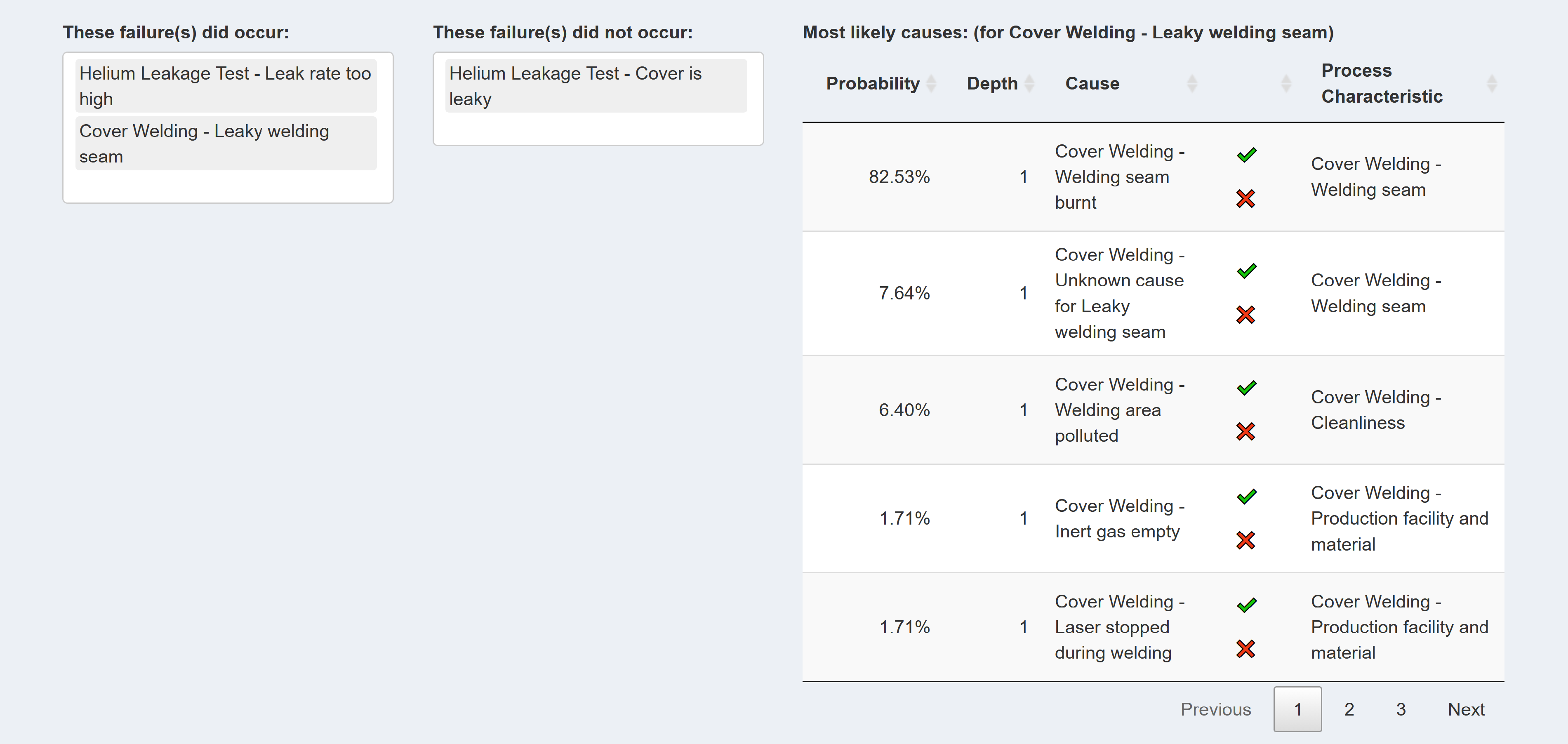}
\vspace{-0.7cm}
\caption{User Interface for RCA.}
\label{fig:UI}
\end{figure}

The user interface with an example of an RCA is shown in Figure \ref{fig:UI}. In this case, the user initially noticed that the leak rate in the helium leakage test was too high. They confirmed that the cover was not leaky and the RCA suggested to check the welding seam. The user noticed that the welding seam was leaky and the RCA now showed the possible causes for this, with the welding seam being burnt ranking highest at an a-posteriori probability of $82.53\%$. The user can interact with the tool by clicking either the check mark or the cross in the suggestions to confirm or dismiss a failure. They can also directly enter this information in the two fields on the left or provide the ID of a specific cell to fill in information on some possible failures automatically based on its recorded data. Moreover, besides the most likely causes, the most likely effects triggered by the given failures can be reviewed for a deeper process understanding. All of the information mentioned is also shown in an interactive graph similar to that in section \ref{chap:RCA}. In future updates, information surveyed in the FMEA on how to detect and rectify individual failures can be shown to provide further assistance.

\section{Conclusion and Prospect} \label{chap:conclusion}
We have presented an FMEA-based method for creating a large-scale knowledge database on possible failures during production from experts, which is transformed into a Bayesian Network under Noisy-OR assumption for Root-Cause Analysis. The consistency of the knowledge is ensured by algorithmically checking for mathematical contradictions and suggesting ways to overcome these. This method was validated in a multi-stage lithium-ion battery prototype production at BMW Group in Munich, Germany.

One key finding is that inconsistencies occur very frequently in the surveyed expert-knowledge, especially in multi-stage production processes with several experts involved. It cannot be distinguished whether these inconsistencies come from the model assumptions -- in particular the Noisy-OR assumption -- or from misjudgments by the experts, or both. In either case, special attention has to be paid to managing these inconsistencies as they carry the risk of rendering the knowledge database useless for further analysis. Our proposed usage of a genetic algorithm to find similar but consistent knowledge bases is a promising first step in enhancing consistency automatically with respect to different levels of certainty on the expert side. Further approaches, such as formulating the task as Constraint Satisfaction Problem (CSP), should be considered with the aim of generating real-time recommendations during knowledge acquisition. 

We found it highly beneficial to transform the rather static FMEA results into an interactive tool for root cause analysis. It makes the combined knowledge of several experienced process experts accessible especially to new and less experienced staff. Bayesian Networks are a scalable and well-interpretable way to represent knowledge-based failure networks mathematically and to perform inference. Once production data becomes available, the expert-based Bayesian Network can be used as a starting point to be advanced by the data. This makes it a supporting tool that can accompany the development from early ramp-up phases to mature series production.

Further research can be conducted to improve the model. As outlined earlier, the currently used Bayesian Network does not take uncertainties of expert statements into account. Once fast inference algorithms for more complex models like Credal Networks become available, it will be beneficial to include this information. Additionally, due to a lack of production data on failures, the model built in this work could not be tested against real observations. Once the production yields data, such a check will serve to quantize the model's performance and make it possible to iteratively refine its parameters.

\section{Acknowledgements}
This research did not receive any specific grant from funding agencies in the public, commercial, or not-for-profit sectors. We would like to thank all process experts that were interviewed for knowledge acquisition.

\bibliography{largeScaleBayesNetworks}

\section*{Appendix: Proof of Leak Probability} \label{chap:leakFormula}
Let $X_i$ be an arbitrary node with existing parents $\text{Pa}(X_i)$ and let $P(L^{(i)} = 1)$ be the (unknown) prior probability of the leak variable $L^{(i)}$ of $X_i$. We can find the value of $P(L^{(i)} = 1)$ that is required to bring the marginal probability of $X_i$ to a predefined value $P(X_i = 0)$ as follows:

\begin{align*}
P\left(X_i = 0\right) = & \sum\limits_{\left(\text{Pa}\left(X_i\right), L^{(i)}\right)} P\left(X_i = 0 \,|\, \text{Pa}\left(X_i\right), L^{(i)}\right) \cdot P\left(\text{Pa}\left(X_i\right), L^{(i)}\right) \\
 		= & \sum\limits_{\left(\text{Pa}\left(X_i\right), L^{(i)}\right)} P\left(X_i = 0 \,|\, \text{Pa}\left(X_i\right), L^{(i)}\right) \cdot P\left(\text{Pa}\left(X_i\right)\right) \cdot P\left(L^{(i)}\right) \\
 		= & \sum\limits_{\left(\text{Pa}\left(X_i\right), L^{(i)} = 0\right)} P\left(X_i = 0 \,|\, \text{Pa}\left(X_i\right), L^{(i)}\right) \cdot P\left(\text{Pa}\left(X_i\right)\right) \cdot P\left(L^{(i)}\right) + \\
 		& \sum\limits_{\left(\text{Pa}\left(X_i\right), L^{(i)} = 1\right)} P\left(X_i = 0 \,|\, \text{Pa}\left(X_i\right), L^{(i)}\right) \cdot P\left(\text{Pa}\left(X_i\right)\right) \cdot P\left(L^{(i)}\right) \\
 		= & \sum\limits_{\text{Pa}\left(X_i\right)} \left(\prod_{j = 1}^J \left(1 - p^{(i)}_j \right)^{X^{(i)}_j}\right) \cdot P\left(\text{Pa}\left(X_i\right)\right) \cdot \left(1 - 1\right)^0 \cdot P(L^{(i)} = 0) + \\
		 & \sum\limits_{\text{Pa}\left(X_i\right)} \left(\prod_{j = 1}^J \left(1 - p^{(i)}_j\right)^{X^{(i)}_j}\right) \cdot P\left(\text{Pa}\left(X_i\right)\right) \cdot \left(1 - 1\right)^1 \cdot P(L^{(i)} = 1) \\
\Leftrightarrow  P(L^{(i)} = 1) = & \, 1 - \frac{P\left(X_i = 0\right)}{\sum\limits_{\text{Pa}\left(X_i\right)} \left(\prod\limits_{j=1}^J \left(1 - p^{(i)}_j\right)^{X^{(i)}_j}\right) \cdot P\left(\text{Pa}\left(X_i\right)\right)}
\end{align*}

\end{document}